\documentclass[twocolumn,twocolappendix]{aastex631}
\usepackage{upgreek}
\usepackage{amsmath,amssymb,amsfonts,bm}
\usepackage{epstopdf}
\usepackage{epsfig}
\usepackage{natbib,caption2}
\usepackage{graphicx}   % Including figure files
\usepackage{float}
\usepackage{longtable}
\usepackage{graphics}
\usepackage{hyperref}
\usepackage{color}
\usepackage{calc}
\usepackage{threeparttable}
\usepackage{ulem}
\usepackage{booktabs}
\usepackage{rotating}
\usepackage{soul}

\newcommand \beq{\begin{equation}}
\newcommand \eeq{\end{equation}}
\newcommand \bey{\begin{eqnarray}}
\newcommand \eey{\end{eqnarray}}

\newcommand \kms{\,{\rm km \, s}^{-1}}

\newcommand{\gsim}{\lower.5ex\hbox{$\; \buildrel > \over \sim \;$}}
\newcommand{\lsim}{\lower.5ex\hbox{$\; \buildrel < \over \sim \;$}}

\newcommand  \siiv  {\ifmmode {\rm Si}\, {\sc iv}\ \else Si\,{\sc iv}\fi}
\newcommand  \SIIV  {\ifmmode {\rm Si}\,{\sc iv}\,\lambda1399 \else Si\,{\sc iv}\,$\lambda1399$\fi}
\newcommand  \civ  {\ifmmode {\rm C}\, {\textsc iv}\ \else C\,{\sc iv}\fi}
\newcommand  \CIV  {\ifmmode {\rm C}\,{\sc iv}\,\lambda1549 \else C\,{\sc iv}\,$\lambda1549$\fi}

\newcommand  \aliii  {\ifmmode {\rm Al}{\textsc{iii}} \else Al\,{\sc iii}\fi}
\newcommand  \ALIII  {\ifmmode {\rm Al}\,{\sc iii]}\,\lambda1854 \else C\,{\sc iii]}\,$\lambda1854$\fi}
\newcommand  \feii {\ifmmode {\rm Fe}\,{\textsc{ii}}\, \else Fe\,{\sc ii}\fi}

\def\mgii{Mg\,{\sc~ii}}
\newcommand  \MGII  {\ifmmode {\rm Mg}\,{\sc ii}\,\lambda2798 \else Mg\,{\sc ii}\,$\lambda2798$\fi}

\newcommand{\mgiiab}{Mg{\sc~ii}\,$\lambda\lambda$2796,2803}

\def\kms{$\rm km\,s^{-1}$}

\def\ergs{${\rm erg\,s^{-1}}$}
\def\Msun{$\rm M_\odot$} %{\rm{M_\odot}}

\shortauthors{Chen et al.}
\begin{document}

\title{The Detection Rate of Associated \mgii\ Absorption Lines in Quasars Depends on Their Radio Emission}

\correspondingauthor{Zhi-Fu Chen}
\email{zhichenfu@126.com}

\author[0000-0003-0639-1148]{Zhi-fu Chen}
\affiliation{School of Mathematics and Physics, Guangxi Minzu University, Nanning 530006, People's Republic of China}

\author[0000-0001-6947-5846]{Luis C. Ho}
\affil{Kavli Institute for Astronomy and Astrophysics, Peking University, Beijing 100871, China}
\affil{Department of Astronomy, School of Physics, Peking University, Beijing 100871, China}

%\author{Zhe-Geng Chen}
%\affiliation{Laboratory for Relativistic Astrophysics, Physical Science and Technology College, Guangxi University, Nanning 530004, People's Republic of China; zhegengc@126.com or zhichenfu@126.com}
%%\affiliation{School of Mathematics and Physics, Guangxi Minzu University, Nanning 530006, People's Republic of China}
%
%\author{Xing-long Peng}
%\affiliation{School of Mathematics and Physics, Guangxi Minzu University, Nanning 530006, People's Republic of China}
%
%\author{Wei-rong Huang}
%\affiliation{School of Mathematics and Physics, Guangxi Minzu University, Nanning 530006, People's Republic of China}

%\collaboration{6}{(AAS Journals Data Editors)}

%% Note that the \and command from previous versions of AASTeX is now
%% depreciated in this version as it is no longer necessary. AASTeX
%% automatically takes care of all commas and "and"s between authors names.

%% AASTeX 6.31 has the new \collaboration and \nocollaboration commands to
%% provide the collaboration states of a group of authors. These commands
%% can be used either before or after the list of corresponding authors. The
%% argument for \collaboration is the collaboration identifier. Authors are
%% encouraged to surround collaboration identifiers with ()s. The
%% \nocollaboration command takes no argument and exists to indicate that
%% the nearby authors are not part of surrounding collaborations.

%% Mark off the abstract in the ``abstract'' environment.
\begin{abstract}
In active galactic nuclei, jet-driven feedback plays a significant role in influencing the properties of gas within their host galaxy and the circumgalactic medium. By combining observations from the Very Large Array Sky Survey, the Faint Images of the Radio Sky at Twenty-cm, the LOFAR Two Metre Sky Survey, and the Sloan Digital Sky Survey, we assembled a sample of 3,141 radio-loud quasars, among which 418 exhibit \mgii\ associated absorption lines in their Sloan spectra. We classify these quasars into evolutionary stages based on their radio spectral shapes. Our analysis reveals that evolved quasars exhibit a significantly higher incidence of \mgii\ associated absorption lines compared to younger sources, particularly among quasars with ``non-peaked'' radio spectra, which show an incidence of \mgii\ associated absorbers approximately 1.7 times greater than that of gigahertz-peaked spectrum sources. This observation can be explained effectively by jet-driven feedback. As quasars age, their jets expand and expel substantial amounts of gas from small scales to larger scales, ultimately reaching the circumgalactic medium. The gas expelled from the inner regions and distributed over larger scales results in a greater coverage fraction of absorbing gas. Consequently, evolved quasars exhibit a higher incidence of \mgii\ absorption lines.
\end{abstract}


\section{Introduction}
It is widely anticipated that absorption features will be present in quasar spectra when the continuum emission from the central regions (such as the accretion disk) of quasars passes through a foreground gas. With regard to the relationship between the absorbing gas medium and the quasar system, these absorption features can be classified broadly into two categories: intervening absorption lines and associated absorption lines. Intervening absorption lines are formed within gas media located far beyond the gravitational potential well of the quasar system \citep[e.g.,][]{1986A&A...155L...8B,2005ApJ...628..637N,2011MNRAS.417..304L,2018ApJ...866...36L,2023ApJS..265...46C}, while associated absorption lines originate from gas physically related to the quasar system \citep[e.g.,][]{2008ApJ...679..239V,2012ApJ...748..131S,2018ApJS..235...11C,2022NatAs...6..339C}. Intervening absorption lines typically have a redshift that is much lower than that of the quasar system ($z_{\rm abs}\ll z_{\rm em}$), whereas associated lines often exhibit redshifts similar to those of the quasar system ($z_{\rm abs}\approx z_{\rm em}$).

For associated absorption lines, the absorbing gas may be found within the outflows or winds produced by quasars \citep[e.g.,][]{1995ApJ...451..498M,2000ApJ...543..686P,2007ApJ...665..990G,2013ApJ...777...56C,2021NatAs...5...13L,2020ApJS..250....3C,2022NatAs...6..339C,2024MNRAS.527.7825D}, as well as in inflows \citep[e.g.,][]{2019Natur.573...83Z,2022A&A...659A.103C}, host galaxies, and the circumgalactic medium (CGM) \citep[e.g.,][]{2008MNRAS.386.2055N,2008MNRAS.388..227W,2016MNRAS.462.3285P,2018MNRAS.481.3865C,2018ApJS..235...11C,2023A&A...673A..89N,2024arXiv240804458H}. As such, associated absorption lines serve as a valuable tool for understanding the intrinsic properties of gas surrounding quasars and the associated dynamics.

It is widely accepted that most massive galaxies host supermassive black holes (SMBHs). These SMBHs are expected to undergo a phase characterized by the accretion of surrounding material through an accretion disk, during which they release a considerable amount of energy. This released energy, manifested in the form of radiation, winds and/or relativistic jets, influences the surrounding environment, including the galaxy's gas distribution, dynamics, ionization structure, morphology, and star formation \citep[e.g.,][]{2009MNRAS.397.1705G,2013MNRAS.433..622M,2018ApJ...866...91C,2022NatAs...6..339C,2024arXiv240918086S}. These processes are collectively referred to as feedback, a key mechanism for reconciling the co-evolution of SMBHs and their host galaxies \citep[e.g.,][]{2012ARA&A..50..455F,2013ARA&A..51..511K,2014ARA&A..52..589H,2018NatAs...2..198H,2022A&ARv..30....6M,2023ApJ...950...16J,2024Natur.632.1009W}.

There are two primary modes of feedback from active galactic nuclei (AGNs): the quasar mode and the radio mode. Quasar-mode feedback is anticipated to be dominant in high-luminosity sources, while radio-mode feedback is typically observed in sources with luminosities significantly below the Eddington limit. In the quasar mode, radiation pressure from the nucleus drives surrounding gas, resulting in the formation of galactic winds. In contrast, in the context of radio-mode feedback, the relativistic jets emitted by AGNs interact with the surrounding gas, thereby regulating the amount of gas in and out of host galaxies.

To comprehend the critical role that AGN feedback plays in galaxy evolution, it is essential to quantify its impact on the surrounding gas throughout the AGN's active lifetime. In radio AGNs, younger systems, along with their corresponding young jets, are generally expected to exhibit jets that operate on small scales, specifically from sub-kpc to kpc distances. These systems typically display optically thick radio spectra that peak in the GHz-MHz frequency range, a phenomenon commonly referred to as gigahertz-peaked spectrum (GPS) or megahertz-peaked spectrum sources \citep[e.g.,][and references therein]{1998PASP..110..493O,2016AN....337....9O,2017ApJ...836..174C,2021A&ARv..29....3O,2022A&A...668A.186S}. The compact size of these jets and their spectra peaking at GHz-MHz frequencies suggest that this youthful phase lasts approximately $10^5$ to $10^6$ years. Due to the limited size of the jets, the interaction cross-section between the jets and the ambient medium is relatively small. Considering both the compact size of the jets and the short timescale associated with GPS sources, it is improbable that the relativistic jets displace significant amounts of surrounding gas from the innermost regions to the more distant outer regions, or to the CGM of galaxies, during the young phase of AGNs.

As AGNs evolve and the jets expand, their radio spectra gradually transition to being optically thin, with the spectral peak shifting to much lower frequencies (often below a few hundred MHz, hereafter referred to as non-peaked AGNs). Consequently, the interaction cross-section between jets and the surrounding medium increases significantly. Active timescales for radio AGNs can extend up to $10^8$ years \citep[e.g.,][]{2013MNRAS.435.3353H,2017MNRAS.469..639H,2017NatAs...1..596M,2019A&A...622A..12H,2023MNRAS.523..620P}. For evolved AGNs, the powerful jets are likely to have expelled substantial amounts of central gas into outer regions or even into the CGM of galaxies. Thus, examining the gas properties of galaxies at various evolutionary stages will enhance understanding of the feedback effects from AGNs.

In this paper, we utilize a large sample of radio quasars to investigate the properties of \mgii\ associated absorption lines among quasars at different evolutionary stages, allowing us to gain deeper insights into the effects of AGN feedback on the surrounding gas.

Throughout this work, we assume a flat $\Lambda$CDM cosmology with parameters $\Omega_m = 0.3$, $\Omega_\Lambda = 0.7$, and $h_0 = 0.7$.

\section{Quasar sample and spectral analysis} \label{sec:sample}
\subsection{Quasar sample}
The Sloan Digital Sky Survey \citep[SDSS;][]{2000AJ....120.1579Y} conducted a detailed mapping of quasars at a resolution of $R\equiv\frac{\lambda}{\Delta \lambda}\approx2000$ across wavelength ranges of $\lambda \approx 3800-9200$ {\AA} or $\lambda \approx 3600-10500$ {\AA} \citep[]{2009ApJS..182..543A,2013AJ....145...10D,2013AJ....146...32S}. The Sixteenth Data Release \citep[DR16Q;][]{2020ApJS..250....8L} provided a catalog of 750,414 spectroscopically confirmed quasars, which exhibit a redshift range of $z\lesssim6$, predominantly concentrated around $z\approx1.5$. To search for \mgiiab\ absorption lines in the SDSS spectra, we initially selected quasars fulfilling the condition $3800/2800<1+z<9000/2800$ (i.e., $1.357 < 1+z < 3.214$, for the SDSS-I/II spectra) or $3600/2800<1+z<10000/2800$ (i.e., $1.286 < 1+z < 3.571$, for the SDSS-III/IV spectra), with $z$ denoting the systemic redshift ($z_{\rm sys}$) as provided by \cite{2022ApJS..263...42W}. This selection process yielded an initial sample of 647,355 quasars.

The Very Large Array Sky Survey \citep[VLASS;][]{2020PASP..132c5001L} mapped the sky across a frequency range of 2--4 GHz (centered at 3000 MHz) with a resolution of approximately $2\farcs5$, resulting in data release for more than 2.6 million objects in Quick Look epoch 2\footnote{Available at {\url{https://cirada.ca/catalogs}}} \citep[][]{2021ApJS..255...30G}. Utilizing the Very Large Telescope (VLA), the Faint Images of the Radio Sky at Twenty-cm (FIRST) observed the North and South Galactic Caps at a frequency centered at 1.4 GHz and with a resolution of approximately $5''$ \citep[][]{1995ApJ...450..559B}, releasing a catalog of 946,432 radio sources\footnote{The FIRST source catalog is available at {\url{https://sundog.stsci.edu/first/catalogs/readme\_14dec17.html}}} \citep[][]{2015ApJ...801...26H}. The LOw-Frequency Array \citep[LOFAR;][]{2017A&A...598A.104S} Two-metre Sky Survey \citep[LoTSS;][]{2013A&A...556A...2V} provides detailed mapping of the northern sky at a resolution of $6''$ over a frequency range of 120--168 MHz (centered at 144 MHz). The second data release \citep[DR2;][]{2022A&A...659A...1S} from LoTSS included data for 4,396,228 objects, of which 4,116,934 are associated with reliable optical data\footnote{{\url{https://cdsarc.cds.unistra.fr/viz-bin/cat/J/A+A/678/A151}}. This radio-optical cross-match catalog is utilized during the cross-matching process between the SDSS DR16Q and LoTSS DR2. \citep[][]{2023A&A...678A.151H}}.

We retained only those sources with $S_{\rm peak}>5\,\sigma$ across all aforementioned radio catalogs, where $S_{\rm peak}$ represents the peak flux density. To investigate the emission characteristics at radio frequencies for the SDSS quasars, we cross-matched our initial sample of 647,355 quasars with the described radio catalogs. Initially, our sample was cross-matched with VLASS within a matching radius of $7''$ \citep[e.g.,][]{2023A&A...674A.198K}, yielding the $\text{SDSS-VLASS}$ sample, which comprises 21,435 quasars. Subsequently, the $\text{SDSS-VLASS}$ sample was cross-matched with LoTSS DR2 within a matching radius of $7''$ \citep[e.g.,][]{2023A&A...674A.198K}, resulting in the $\text{SDSS-VLASS-LoTSS}$ sample consisting of 10,483 quasars. Finally, the $\text{SDSS-VLASS-LoTSS}$ sample was cross-matched with FIRST within a matching radius of $5''$ \citep[e.g.,][]{2011ApJS..194...45S}, producing the $\text{SDSS-VLASS-LoTSS-FIRST}$ sample containing 4,045 quasars. To ensure an effective survey of \mgii\ absorption lines, we also excluded those quasars with a median signal-to-noise ratio (S/N) less than 5 (median $\rm S/N \leq 5~pixel^{-1}$), where the median S/N was computed using spectral data from the red wing ranging from 5000 \kms\ to 20000 \kms\ of the \mgii\ emission line. Following this filtering process, our parent sample consists of 3,726 quasars, whose properties across the optical and radio bands are presented in Table \textcolor{blue}{\ref{tab:Table1}}.

%XX I edited many font issues below. You still need to change minus signs to $-$
%
\begin{table*}[htbp]
%\begin{sidewaystable*}
\tiny
\caption{The Properties of the Sample}\centering \tabcolsep 0.7mm \scriptsize%\footnotesize%\small%\scriptsize
\label{tab:Table1}
\begin{tabular}{cccccccccccccccc}
\hline\hline\noalign{\smallskip}
SDSS name &$z_{\rm em}$ &  $\log L_{\rm 3000}$  & $\log M_{\rm BH}$ &$S_{\rm peak}^{\rm VLASS}$&$S_{\rm peak}^{\rm FIRST}$&$S_{\rm peak}^{\rm LoTSS}$& $R$&$\alpha^{\rm 144}_{\rm 1400}$&$\alpha^{\rm 1400}_{\rm 3000}$&$z_{\rm abs}$&$\rm EW^{\lambda2796}$&$\rm EW^{\lambda2803}$ &$\upsilon_r$ \\
&&(\ergs)&(\Msun) &(mJy)&(mJy)&(mJy)& &&&&(\AA)&(\AA) &(\kms)\\
\hline\noalign{\smallskip}
072400.39+354950.0	&	1.657 &	46.04 	&	8.94 &	1.0$\pm$0.1 	&	1.7$\pm$0.2 	&	4.5$\pm$0.1 	&	1.62 	&	-0.44$\pm$0.09 	&	-0.55$\pm$0.15 	&	$-$9999	&	$-$9999			&	$-$9999			&	$-$9999	\\
072506.15+262455.7	&	2.147 &	45.94 	&	8.86 &	69.9$\pm$0.3 	&	106.7$\pm$0.1 	&	287.8$\pm$0.5 	&	4.44 	&	-0.44$\pm$0.00 	&	-0.37$\pm$0.00 	&	$-$9999	&	$-$9999			&	$-$9999			&	$-$9999	\\
072623.80+395539.9	&	1.563 &	45.63 	&	8.93 &	0.9$\pm$0.1 	&	1.5$\pm$0.1 	&	3.0$\pm$0.1 	&	1.89 	&	-0.30$\pm$0.10 	&	-0.54$\pm$0.16 	&	$-$9999	&	$-$9999			&	$-$9999			&	$-$9999	\\
072736.72+355315.5	&	1.436 &	45.45 	&	9.24 &	5.8$\pm$0.2 	&	4.2$\pm$0.1 	&	22.5$\pm$0.1 	&	2.88 	&	-0.74$\pm$0.04 	&	0.61$\pm$0.04 	&	$-$9999	&	$-$9999			&	$-$9999			&	$-$9999	\\
072841.99+353525.6	&	1.444 &	45.17 	&	9.54 &	1.1$\pm$0.1 	&	1.9$\pm$0.2 	&	2.2$\pm$0.1 	&	2.29 	&	-0.05$\pm$0.09 	&	-0.56$\pm$0.13 	&	$-$9999	&	$-$9999			&	$-$9999			&	$-$9999	\\
072843.03+370835.1	&	1.380 &	45.80 	&	9.66 &	1.7$\pm$0.1 	&	1.1$\pm$0.1 	&	1.5$\pm$0.1 	&	1.74 	&	-0.15$\pm$0.15 	&	0.82$\pm$0.15 	&	1.342	&	0.28$\pm$0.09 	&	0.43$\pm$0.14 	&	4829	\\
072843.03+370835.1	&	1.380 &	45.80 	&	9.66 &	1.7$\pm$0.1 	&	1.1$\pm$0.1 	&	1.5$\pm$0.1 	&	1.74 	&	-0.15$\pm$0.15 	&	0.82$\pm$0.15 	&	1.404	&	1.27$\pm$0.11 	&	0.91$\pm$0.10 	&	$-$3097	\\
072909.14+324418.6	&	2.204 &	45.90 	&	8.71 &	4.0$\pm$0.1 	&	4.9$\pm$0.1 	&	14.9$\pm$0.1 	&	2.79 	&	-0.49$\pm$0.03 	&	-0.08$\pm$0.04 	&	$-$9999	&	$-$9999			&	$-$9999			&	$-$9999	\\
072928.48+252451.8	&	2.310 &	46.74 	&	9.32 &	31.7$\pm$0.2 	&	6.8$\pm$0.1 	&	5.1$\pm$0.2 	&	2.63 	&	0.13$\pm$0.05 	&	2.20$\pm$0.02 	&	$-$9999	&	$-$9999			&	$-$9999			&	$-$9999	\\
073023.70+324836.1	&	1.252 &	45.74 	&	9.66 &	5.4$\pm$0.1 	&	4.5$\pm$0.1 	&	4.4$\pm$0.1 	&	2.04 	&	0.01$\pm$0.04 	&	0.41$\pm$0.04 	&	$-$9999	&	$-$9999			&	$-$9999			&	$-$9999	\\
073211.10+321245.8	&	1.549 &	45.44 	&	9.25 &	26.2$\pm$0.1 	&	18.6$\pm$0.1 	&	26.6$\pm$0.2 	&	3.20 	&	-0.16$\pm$0.01 	&	0.63$\pm$0.01 	&	1.532	&	0.48$\pm$0.08 	&	0.31$\pm$0.13 	&	2019	\\
......&......&......&......&......&......&......&......&......&......&......&......&......&......\\
\noalign{\smallskip}
\hline\hline\noalign{\smallskip}
\end{tabular}
\begin{flushleft}
\small
The radio-loudness parameter $R = \log~L_{\rm 1.4\,GHz}/L_i$. The radio spectral index $\alpha^{\rm 144}_{\rm 1400}$ is defined between 144 MHz and 1400 MHz, and $\alpha^{\rm 1400}_{\rm 3000}$ is defined between 1400 and 3000 MHz. A value of $-$9999 indicates the absence of detected \mgii\ absorption. The velocity offset of the \mgii\ absorption line is computed via $\upsilon_{r}=c\times\frac{(1+z_{\rm em})^2-(1+z_{\rm abs})^2}{(1+z_{\rm em})^2+(1+z_{\rm abs})^2}$, where $c$ is the speed of light. The quasars listed in Table are just a small selection of the full sample. See the online table for the full sample.\\
(This table is available in its entirety in machine-readable form.)
\end{flushleft}
\end{table*}
%\end{sidewaystable*}

\begin{figure}[ht!]
\includegraphics[width=0.48\textwidth]{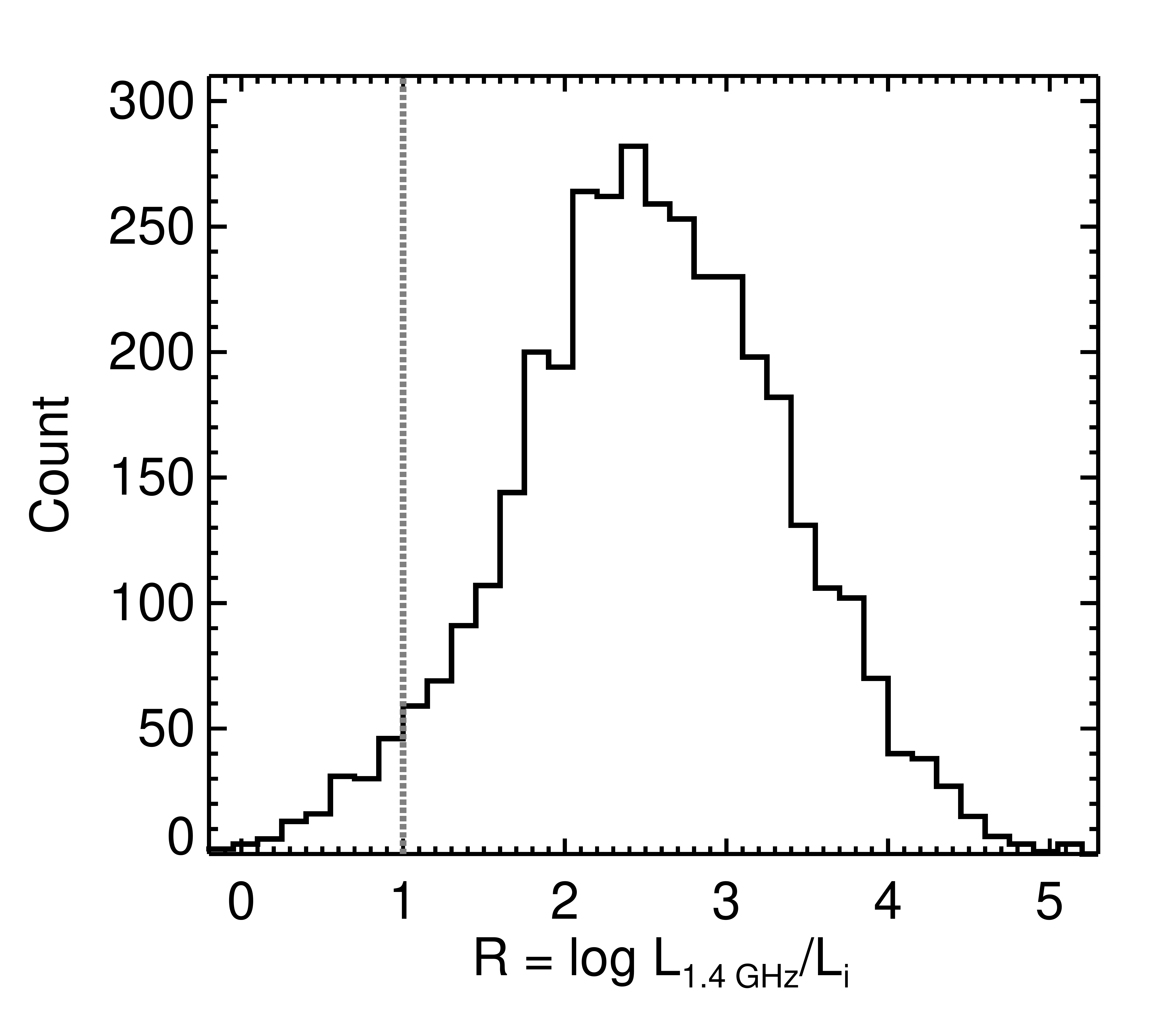}
\caption{The radio-loudness for all quasars included in our parent sample. The gray dashed line indicates the boundary distinguishing radio-quiet ($R<1$) from radio-loud (\( R \ge 1 \)) quasars. Our final sample is comprised solely of radio-loud sources.}
\label{fig:loudness}
\end{figure}

\begin{figure}[ht!]
\includegraphics[width=0.49\textwidth]{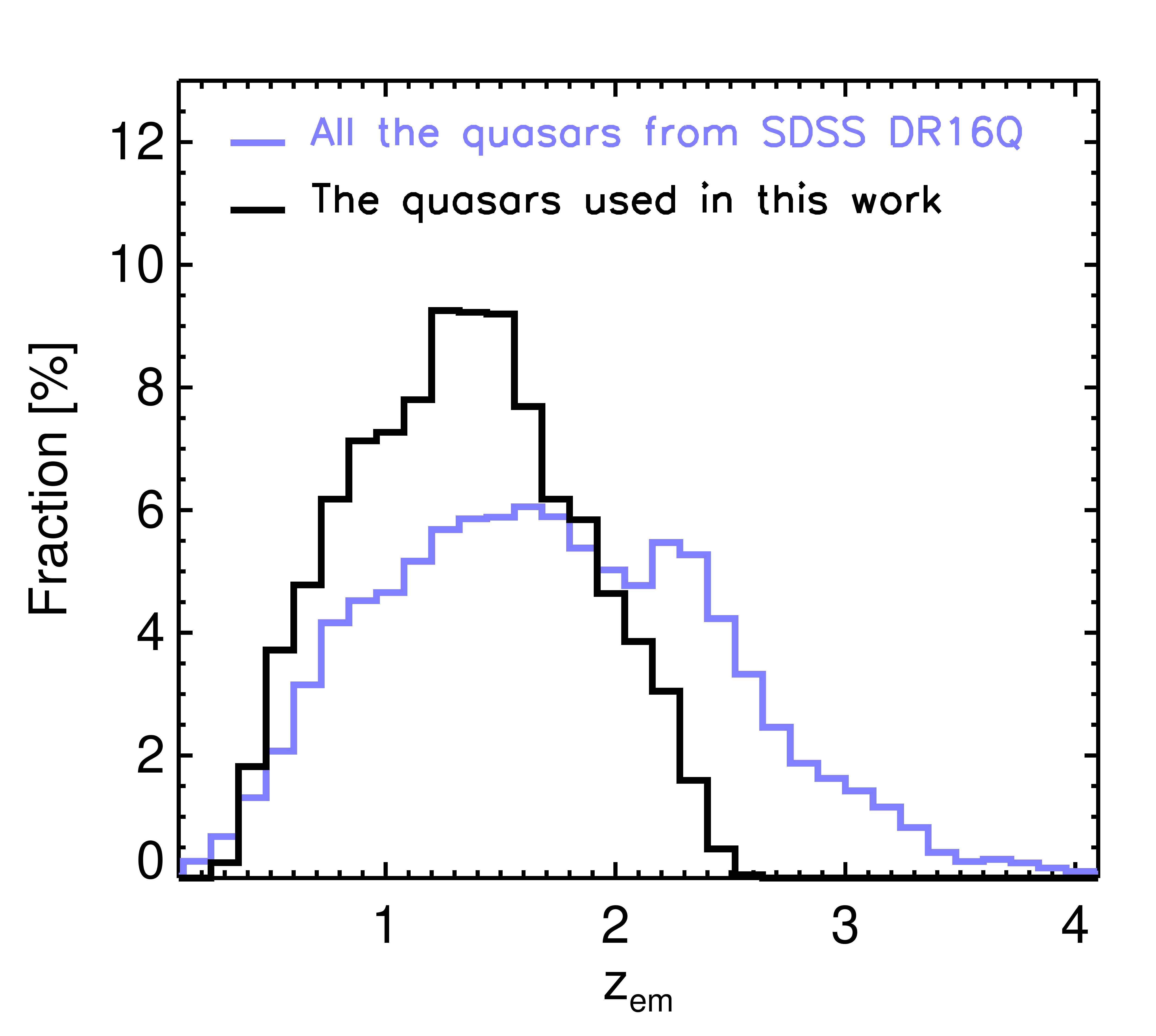}
\caption{The redshift distribution of quasars. The light-blue line is for all the quasars from the SDSS DR16Q. The black line is for the quasars included in our final sample. The Y-axis has been normalized by the total numbers of quasars included in corresponding catalogs.}
\label{fig:ze}
\end{figure}

The ratio of luminosities at 1400 MHz ($L_{\rm 1.4\,GHz}$) to the $i$-band luminosity ($L_{i}$) is frequently employed to quantify radio-loudness in quasars. A commonly used threshold is $R=\log L_{\rm 1.4\,GHz}/L_{i}=1$, which separates radio-quiet quasars (with $R<1$) from radio-loud quasars (with $R\ge1$) \citep[e.g.,][]{2012ApJ...759...30B}. For our parent quasar sample, we calculate the radio spectral index between 3000 MHz (VLASS) and 144 MHz (LoTSS) using the convention $S_{\rm\nu} \propto \nu^{\alpha} $:
\begin{equation}\label{eq:index144_3000}
  \alpha^{\rm 144}_{\rm 3000} = \frac{\log S_{\rm\nu}^{\rm VLASS} - \log S_{\rm\nu}^{\rm LoTSS}}{\log \nu_{\rm VLASS} - \log \nu_{\rm LoTSS}},
\end{equation}
where $S_{\rm\nu}^{\rm VLASS}$ and $S_{\rm\nu}^{\rm LoTSS}$ denote the peak flux densities at 3000 and 144 MHz, respectively. Utilizing the radio spectral index $\alpha_{3000}^{144}$, we derive $L_{\rm 1.4\,GHz}$ from the flux density at 144 MHz. The $i$-band luminosity $L_{i}$ is calculated from the absolute $i$-band magnitude at $z=0$ ($M_{i}^{z=0}$) using the equation
\begin{equation}\label{eq:mi}
  M_{i}^{z=0} = M_{i}^{z=2} + 2.5 \,(1 + \alpha_{\rm opt} \times \log(1 + z)),
\end{equation}
where $M_{i}^{\rm z=2}$ is the absolute magnitude in the $i$ band derived from DR16Q \cite[][]{2020ApJS..250....8L} and $k$-corrected to $z = 2$, and $\alpha_{\rm opt}=-0.5$ is the optical spectral index. We compute the radio-loudness for our parent quasar sample based on the available values of $L_{\rm 1.4\,GHz}$ and $L_{i}$. The results of this computation are presented in Table \textcolor{blue}{\ref{tab:Table1}} and are graphically represented in Figure \textcolor{blue}{\ref{fig:loudness}}. Our final sample consists of 3,577 quasars with $R\ge1$ selected from the parent sample. Figure \textcolor{blue}{\ref{fig:ze}} illustrates the redshift distribution of the quasars included in our final sample, all of which fall within the range $0.2877\le z_{\rm em}\le2.5604$.

%XX X-axis, make fonts consistent with main text (e.g., use italics to mimic $$).  Log --> log
%

%XX X-axis, make fonts consistent with main text (e.g., use italics to mimic $$).
%   Don't use gray, which is hard to use. Use different color? If you adopt this, then edit the captions accordingly.
%

\subsection{Spectral analysis}
\begin{figure*}[ht!]
\centering
\includegraphics[width=0.65\textwidth]{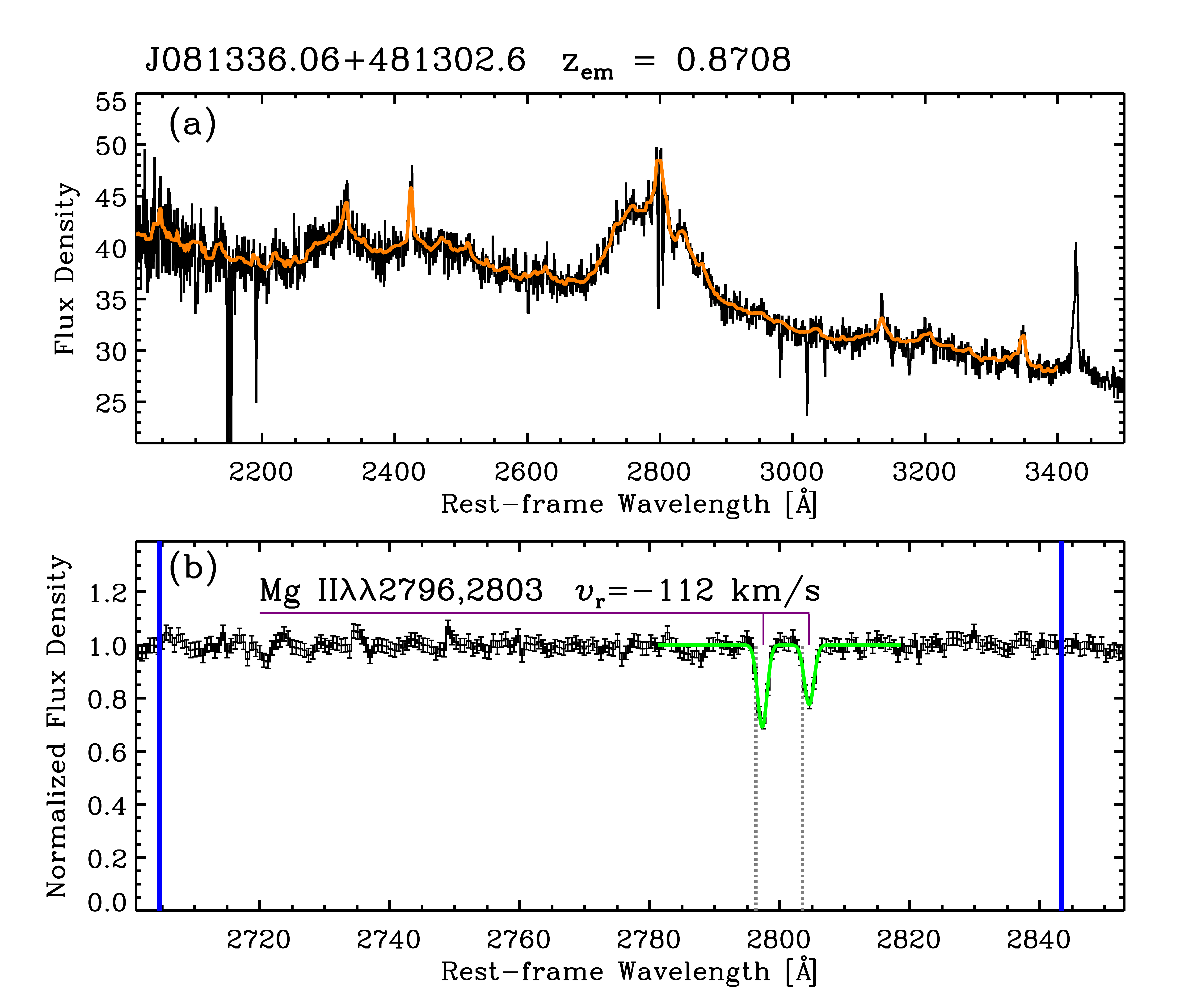}
\includegraphics[width=0.65\textwidth]{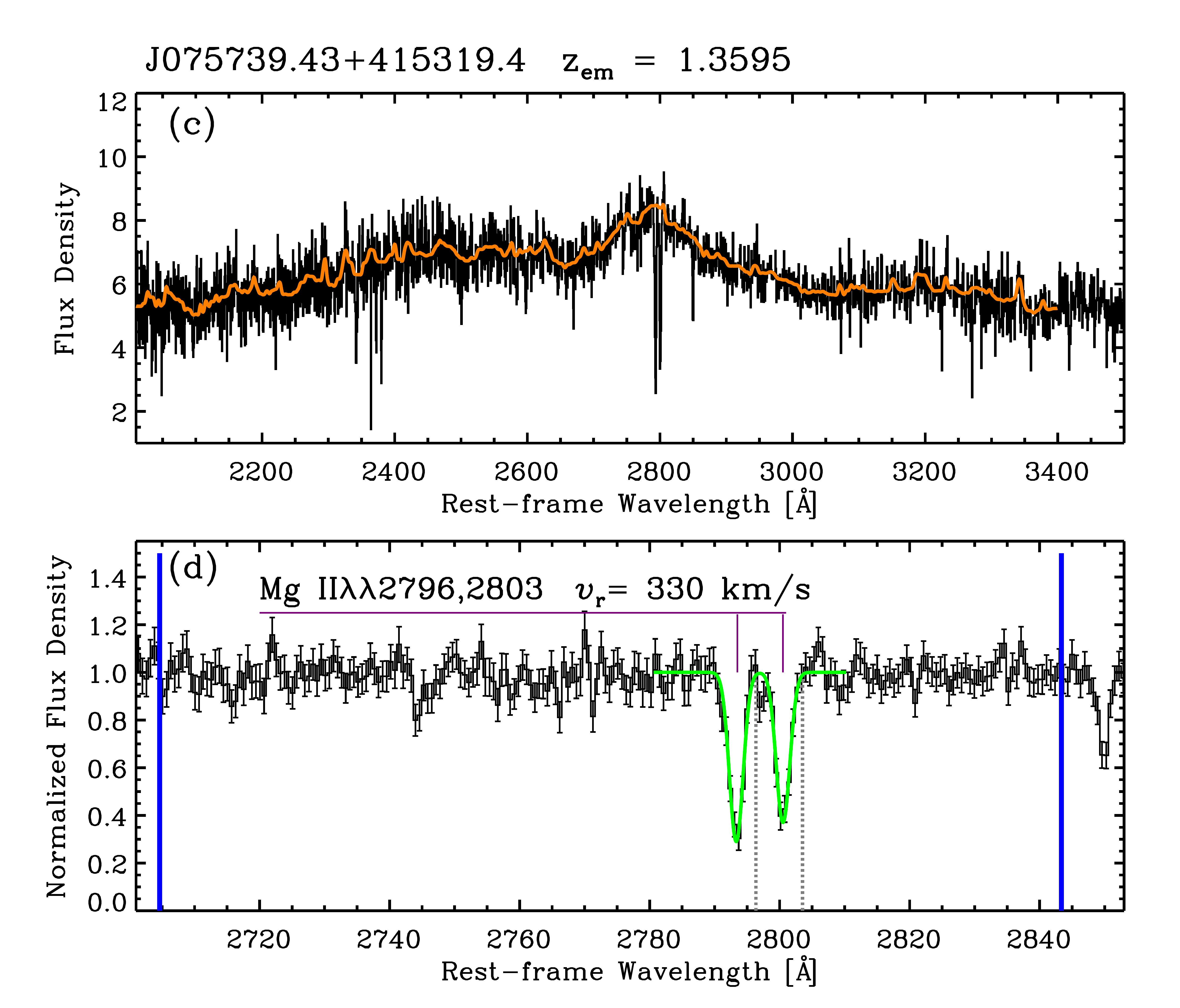}
\caption{The \mgii\ absorption lines imprinted in the spectra of quasar SDSS~$\rm J081336.06+481302.6$ and SDSS~$\rm J075739.43+415319.4$. Panels (a) and (c): the quasar spectra overlaid with the pseudo-continuum fits (the orange solid lines). Panels (b) and (d): the quasar flux and associated flux uncertainty (the black solid lines and error bars) normalized by the pseudo-continuum fits. The blue solid lines delineate the boundaries of the spectral regions used to search for \mgii\ associated absorption lines. The green solid lines correspond to the Gaussian function fits for the \mgii\ absorption lines with $\upsilon_{r} = -112$ \kms\ (panel b) or $\upsilon_{r} = 300$ \kms\ (panel d). The gray dotted lines indicate the positions of the \mgii\ absorption lines in the quasar's rest frame.}
\label{fig:sample_fig}
\end{figure*}

To investigate the \mgii\ absorption lines that are physically associated with the radio-loud quasars in our final sample, we model the SDSS spectra using methodologies consistent with our previous studies \citep[e.g.,][]{2018ApJS..235...11C,2023ApJS..265...46C}. The initial step involves fitting a pseudo-continuum for each quasar using a combination of cubic spline functions and multi-Gaussian functions in an iterative manner to represent the underlying continuum and emission lines, respectively. Figure \textcolor{blue}{\ref{fig:sample_fig}} illustrates the fitted pseudo-continuum, indicated by the orange solid line. In the second step, we search for candidate \mgii\ absorption lines in the spectra that have been normalized by the pseudo-continuum fits (refer to Figure \textcolor{blue}{\ref{fig:sample_fig}}b). Although a few outflow \mgii\ absorption lines exhibit velocities exceeding 10,000 \kms\ \citep[e.g.,][]{2012ApJ...754...38C,2013ApJ...777...56C,2018MNRAS.473.5154M,2020ApJS..250....3C}, the occurrence of ultra high-speed \mgii\ absorption lines is expected to be minimal. Consequently, the vast majority of \mgii\ associated absorption lines are anticipated to have velocities less than 6,000 \kms\ \citep[e.g.,][]{2018ApJS..235...11C}. Therefore, we identify the \mgii\ absorption features within the spectral range from the blue wing at 10,000 \kms\ to the red wing at $-$5,000 \kms\ relative to the \mgii\ emission line (as depicted by the blue solid lines in Figure \textcolor{blue}{\ref{fig:sample_fig}}). Each candidate of the \mgii\ absorption doublet is modeled using a combination of two Gaussian functions, represented by the green solid lines in Figure \textcolor{blue}{\ref{fig:sample_fig}}. The identification process of \mgii\ absorption lines is guided by the separation of the \mgii\ doublet and involves a visual inspection of each line individually. The equivalent width (EW) of the absorption lines is calculated by integrating the Gaussian function fits. The integration of the flux uncertainty, normalized by the pseudo-continuum fit within $\pm3\,\sigma$, yields the error in the equivalent width ($\sigma_{\rm EW}$), where $\sigma$ corresponds to the Gaussian function fits. We retain only those \mgii\ absorption lines that satisfy the criteria $\rm EW^{\rm \lambda2796} \ge 0.1$ \AA, $\rm EW^{\rm \lambda2803} \ge 0.1$ \AA, $\rm EW^{\rm \lambda2796} \ge 2\,\sigma_{\rm EW}$, and $\rm EW^{\rm \lambda2803} \ge 2\,\sigma_{\rm EW}$. As a result, we identify a total of 597 \mgii\ absorption lines from the spectra of 3,577 radio-loud quasars. These absorption lines were detected in 522 distinct quasars. The relevant absorption-line parameters are presented in Table \textcolor{blue}{\ref{tab:Table1}}.

%XX There are many formatting and labelling issues. I have no time to note them all. Just remember, everything in figures and tables should be the same as the main text.  X-axis: Quasar Rest Frame --> Rest-frame Wavelength
%
%    Try to put the labels (a), (b), (c), (d) inside each panel.
%%

\section{Results and Discussions} \label{sec:DISCUS}
\begin{figure}[ht!]\centering
\includegraphics[width=0.49\textwidth]{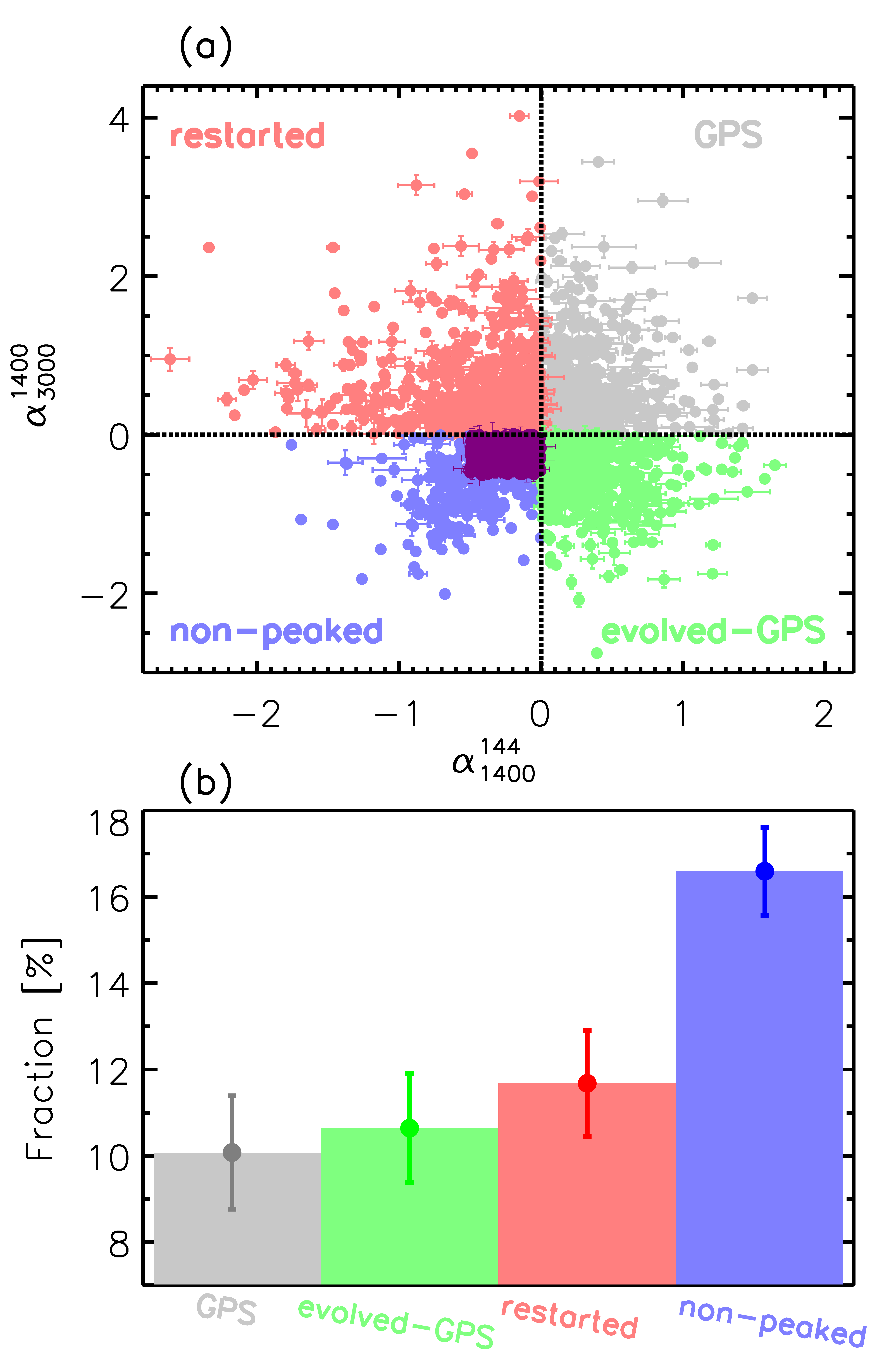}
\caption{(a): The distribution of spectral indices for the 3,577 quasars included in our final sample. The designations ``GPS'', ``evolved-GPS'', ``restarted'', and ``non-peaked'' are represented by light-gray, light-green, light-red, and light-blue symbols, respectively. The GPS quasars are situated in the earliest evolutionary stage, while the non-peaked quasars represent the most advanced evolutionary stage. The purple symbols denote quasars with $-0.5 \le \alpha^{\rm 144}_{\rm 1400} < 0$ and $-0.5 \le \alpha^{\rm 1400}_{\rm 3000} < 0$, henceforth referred to as flat-spectrum sources. It is important to note that the non-peaked group does not include flat-spectrum quasars. (b): The distribution of detection rates for \mgii\ associated absorption lines. In both panels (a) and (b), symbols of the same color represent the same category of quasars.}
\label{fig:fr_abs}
\end{figure}

The radio spectral shape is frequently regarded as a reliable means of tracing the evolutionary stages of quasars \citep[e.g.,][]{1994A&A...285...27K,2017NatAs...1..596M,2020A&A...638A..29B,2021Galax...9...88M,2023A&A...674A.198K,2024ApJ...974..277C,2024A&A...690A.140K}. To characterize the radio spectral shapes of quasars, we calculate the spectral indices between 3000 MHz (VLASS) and 1400 MHz (FIRST) using
\begin{equation}\label{eq:index1400_3000}
  \alpha^{\rm 1400}_{\rm 3000} = \frac{\log S_{\rm\nu}^{\rm VLASS} - \log S_{\rm\nu}^{\rm FIRST}}{\log \nu_{\rm VLASS} - \log \nu_{\rm FIRST}},
\end{equation}
and between 1400 MHz (FIRST) and 144 MHz (LoTSS) following
\begin{equation}\label{eq:index144_1400}
  \alpha^{\rm 144}_{\rm 1400} = \frac{\log S_{\rm\nu}^{\rm FIRST} - \log S_{\rm\nu}^{\rm LoTSS}}{\log \nu_{\rm FIRST} - \log \nu_{\rm LoTSS}},
\end{equation}
where $S_{\rm\nu}^{\rm VLASS}$, $S_{\rm\nu}^{\rm FIRST}$, and $S_{\rm\nu}^{\rm LoTSS}$ represent the peak flux densities at 3000, 1400, and 144 MHz, respectively. The results are illustrated in Figure \textcolor{blue}{\ref{fig:fr_abs}}. We identify the radio spectral shapes of quasars according to the positions of the sources in Figure \textcolor{blue}{\ref{fig:fr_abs}}.

Quasars with $\alpha^{\rm 144}_{\rm 1400}>0$ and $\alpha^{\rm 1400}_{\rm 3000}>0$ exhibit a radio spectrum with a peaked frequency $\nu_{p}>3000$ MHz (indicated by the light-gray symbols in Figure \textcolor{blue}{\ref{fig:fr_abs}}). These quasars are in the earliest stage of their evolution and are referred to as GPS  sources. As these quasars evolve, the spectral peak shifts to a lower frequency range; however, they still retain a spectral shape with $\alpha^{\rm 144}_{\rm 1400}>0$ (represented by the light-green symbols in Figure \textcolor{blue}{\ref{fig:fr_abs}}). These sources are categorized as ``evolved-GPS'' quasars. With continued evolution, the spectral peaks of quasars shift to significantly lower frequency ranges, characterized by $\alpha^{\rm 144}_{\rm 1400}<0$ and $\alpha^{\rm 1400}_{\rm 3000}<0$ (denoted by the light-blue symbols in Figure \textcolor{blue}{\ref{fig:fr_abs}}). These quasars represent the oldest sources included in our sample and are classified as ``non-peaked'' sources. Additionally, quasars exhibiting $\alpha^{\rm 144}_{\rm 1400}<0$ and $\alpha^{\rm 1400}_{\rm 3000}>0$ display complex radio spectra (marked by the light-red symbols in Figure \textcolor{blue}{\ref{fig:fr_abs}}), and they may be in a phase of periodic activity and are identified as ``restarted'' sources. Note that some quasars possess spectra with $-0.5\le\alpha^{\rm 144}_{\rm 1400}<0$ and $-0.5\le\alpha^{\rm 1400}_{\rm 3000}<0$, which are designated as ``flat-spectrum'' sources. The spectral index of these sources could be affected by relativistic beaming. Thus, these flat-spectrum quasars are not included in the non-peaked category. Based on these definitions, our final quasar sample comprises 526 GPS, 592 evolved-GPS, 685 restarted, and 1338 non-peaked quasars. Among these 3141 quasars, 418 sources exhibit detected \mgii\ associated absorption lines.

\begin{figure}[ht!]
\includegraphics[width=0.48\textwidth]{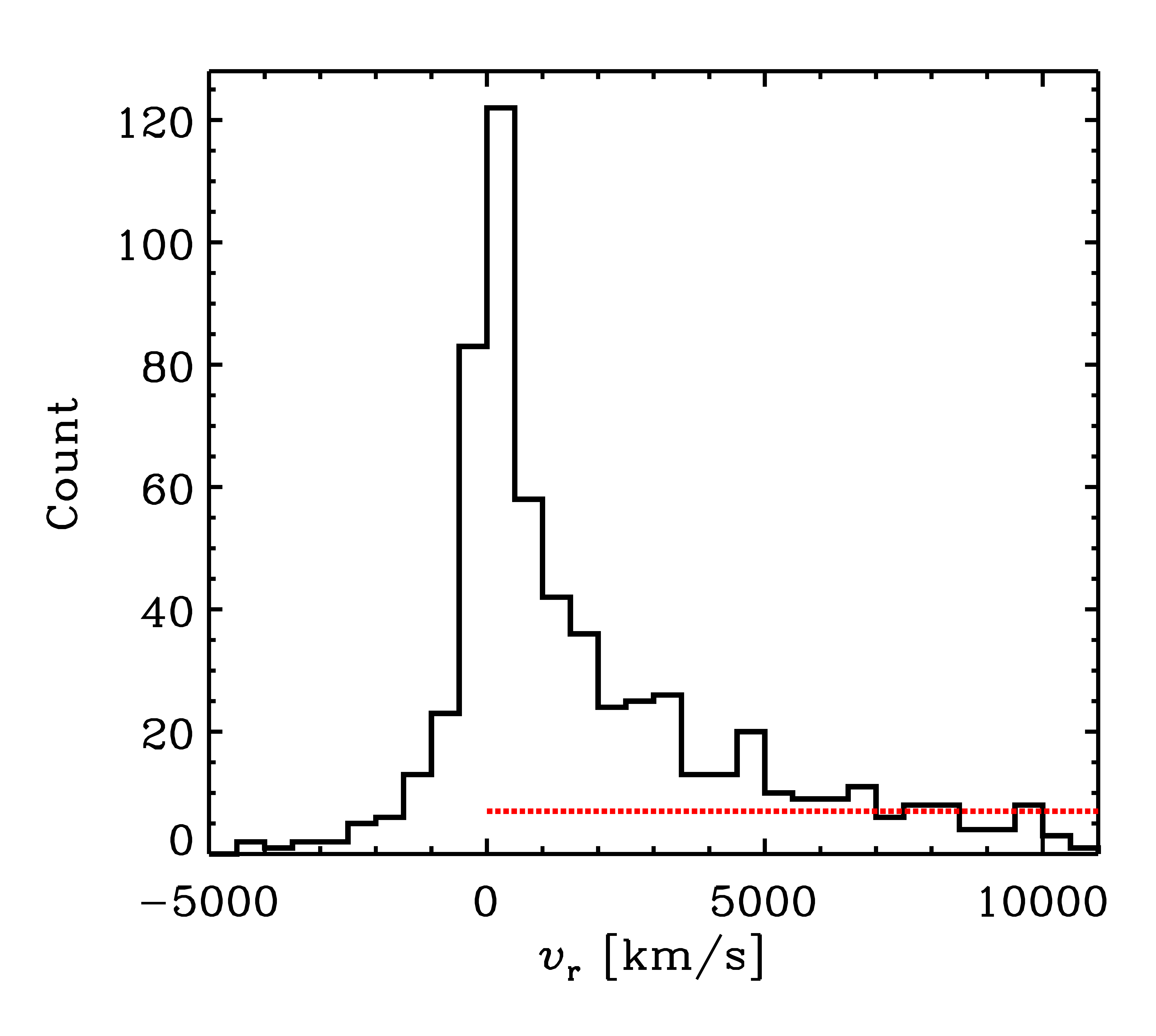}
\caption{The distribution of velocity offset ($\upsilon_{r}$) for the \mgii\ absorption lines detected in quasar spectra. Red-dash line is the mean count of \mgii\ absorption lines in velocity range of 5000 \kms\ $<\upsilon_{r}<$ 10000 \kms.}
\label{fig:vr}
\end{figure}

This study utilizes \mgii\ associated absorption lines to investigate the properties of gas within the gravitational potential well of quasar systems. According to \cite{2018ApJS..235...11C}, a velocity offset\footnote{The velocity offset of \mgii\ absorption line is computed via $\upsilon_{r}=c\times\frac{(1+z_{\rm em})^2-(1+z_{\rm abs})^2}{(1+z_{\rm em})^2+(1+z_{\rm abs})^2}$, where $c$ is the speed of light.} of $\upsilon_{r}<6000$ \kms\ serves as a reliable boundary for estimating \mgii\ associated absorption lines, even though some high-speed outflows may exceed this threshold. Figure \textcolor{blue}{\ref{fig:vr}} illustrates the velocity offsets for \mgii\ absorption lines detected in GPS, evolved-GPS, restarted, and non-peaked quasars. This figure clearly indicates that the number of \mgii\ absorption lines with $\upsilon_{r}>5000$ \kms remains relatively consistent, which aligns with the velocity boundary established by \cite{2018ApJS..235...11C}. This observation implies that \mgii\ absorption lines with $\upsilon_{r}>5000$ \kms\ are likely formed within intervening clouds, while the majority of \mgii\ associated absorption lines have a velocity offset of $\upsilon_{r}<5000$ \kms. Consequently, we restrict our analysis to \mgii\ associated absorption lines with velocity offsets of $\upsilon_{r}<5000$ \kms.

\begin{figure*}[ht!]
\includegraphics[width=0.48\textwidth]{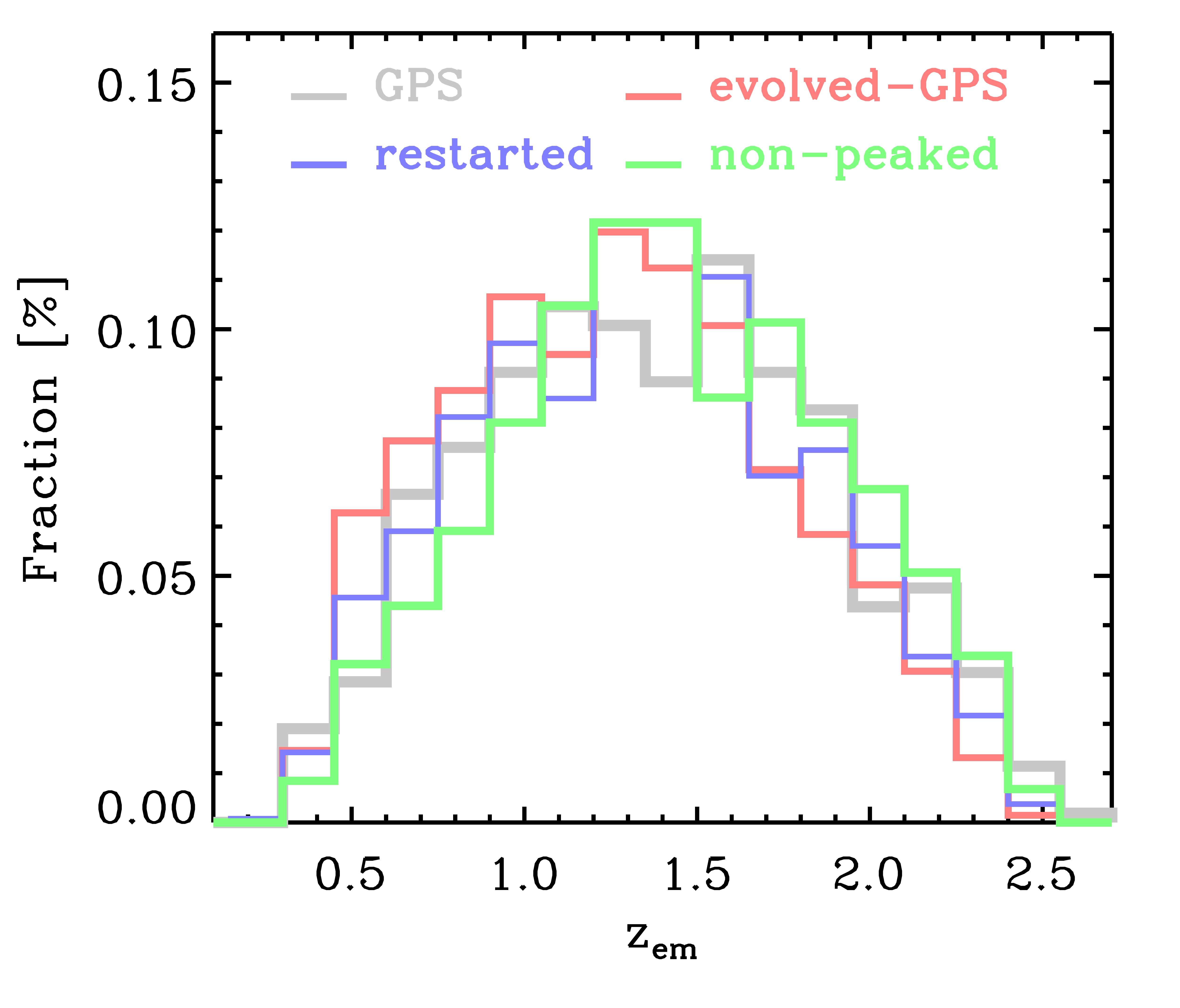}
\includegraphics[width=0.48\textwidth]{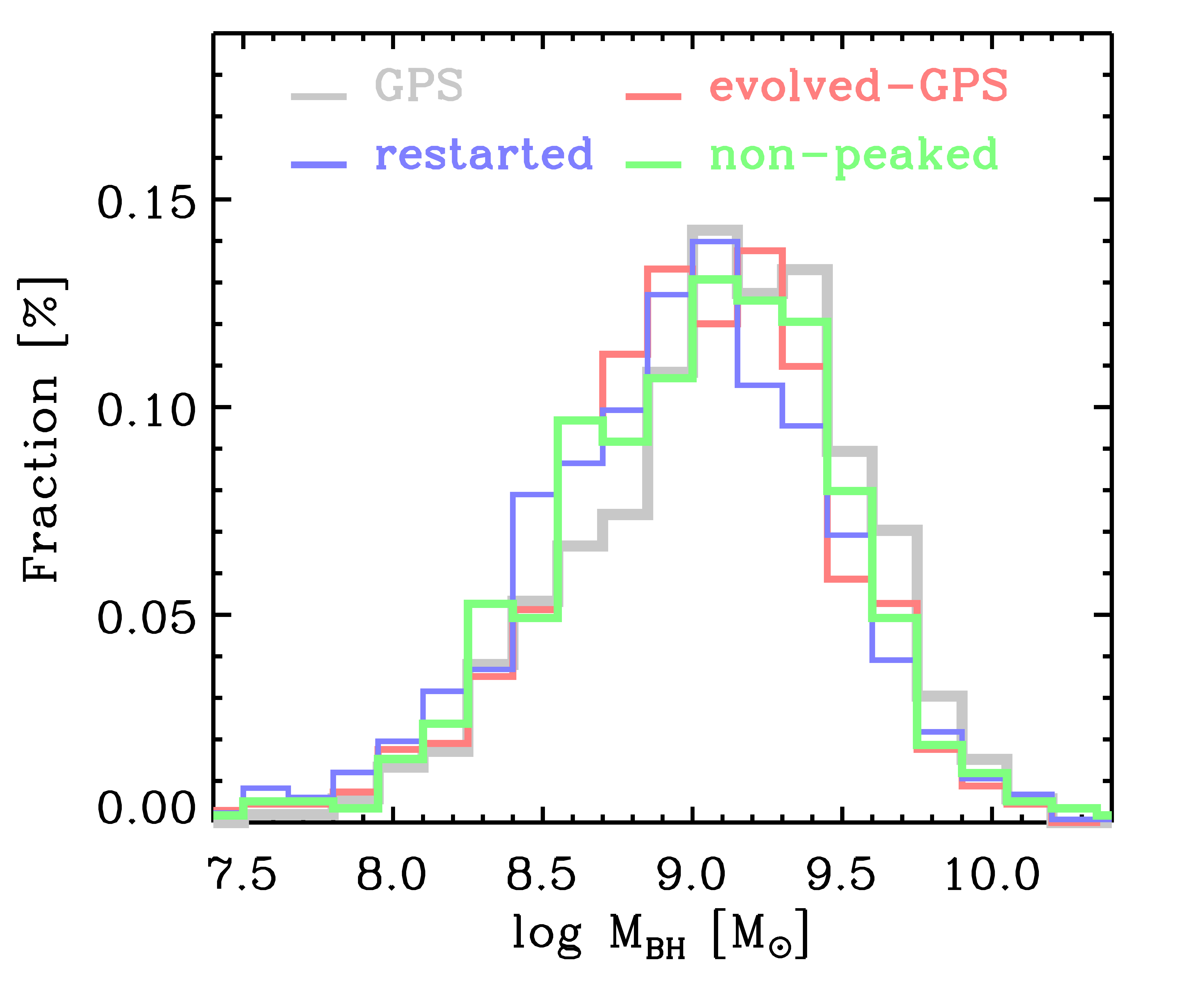}
\caption{Left panel: quasar redshift distributions. Right panel: black hole mass distributions. The light-gray, light-green, light-red, and the light-blue lines represent the GPS, evolved-GPS, restarted, and non-peaked quasars respectively.}
\label{fig:ze_mbh_bin}
\end{figure*}

For each group of quasars, we calculate the incident rate of \mgii\ associated absorption lines,
\begin{equation}\label{eq:fc}
  f_{c} = \frac{N_{\rm abs}}{N_{\rm QSO}},
\end{equation}
where $N_{\rm abs}$ represents the number of detected \mgii\ associated absorption lines, and $N_{\rm QSO}$ denotes the number of quasars utilized in the search for \mgii\ associated absorption lines. Based on the definition of $f_{c}$, this parameter reflects the coverage fraction of the absorbing cloud relative to the background continuum emission regions. Figure \textcolor{blue}{\ref{fig:fr_abs}} shows the resultant values of $f_{c}$, indicating an overall trend of increasing $f_{c}$ as quasars progress from the evolutionary stage. Notably, non-peaked quasars exhibit a significantly higher incidence rate of \mgii\ associated absorption compared to other quasar types. For each quasar category, Figure \textcolor{blue}{\ref{fig:ze_mbh_bin}} gives the distributions of quasar redshifts and black hole masses, demonstrating that each group hosts a comparable distribution of redshift and black hole mass. Here the black hole mass are directly taken from the fiducial single-epoch black hole mass provided by \cite{2022ApJS..263...42W}. These findings suggest that the markedly higher incidence rate of \mgii\ associated absorption in non-peaked quasars is not correlated with quasar redshift or black hole mass.

%XX It is better to include a legend. A good figure is one that can be understood without reading the captions. Apply this philosophy to all the figures
%
%\begin{figure}[ht!]
%\includegraphics[width=0.48\textwidth]{mbh_z.pdf}
%\caption{Black hole mass vs quasar redshift. The light-gray circles, light-green squares, light-red stars, and light-blue triangles represent the ``GPS'', ``evolved-GPS'', ``restarted'', and ``non-peaked'' quasars, respectively.}
%\label{fig:ze_mbh_bin}
%\end{figure}

\begin{figure}[ht!]
\includegraphics[width=0.48\textwidth]{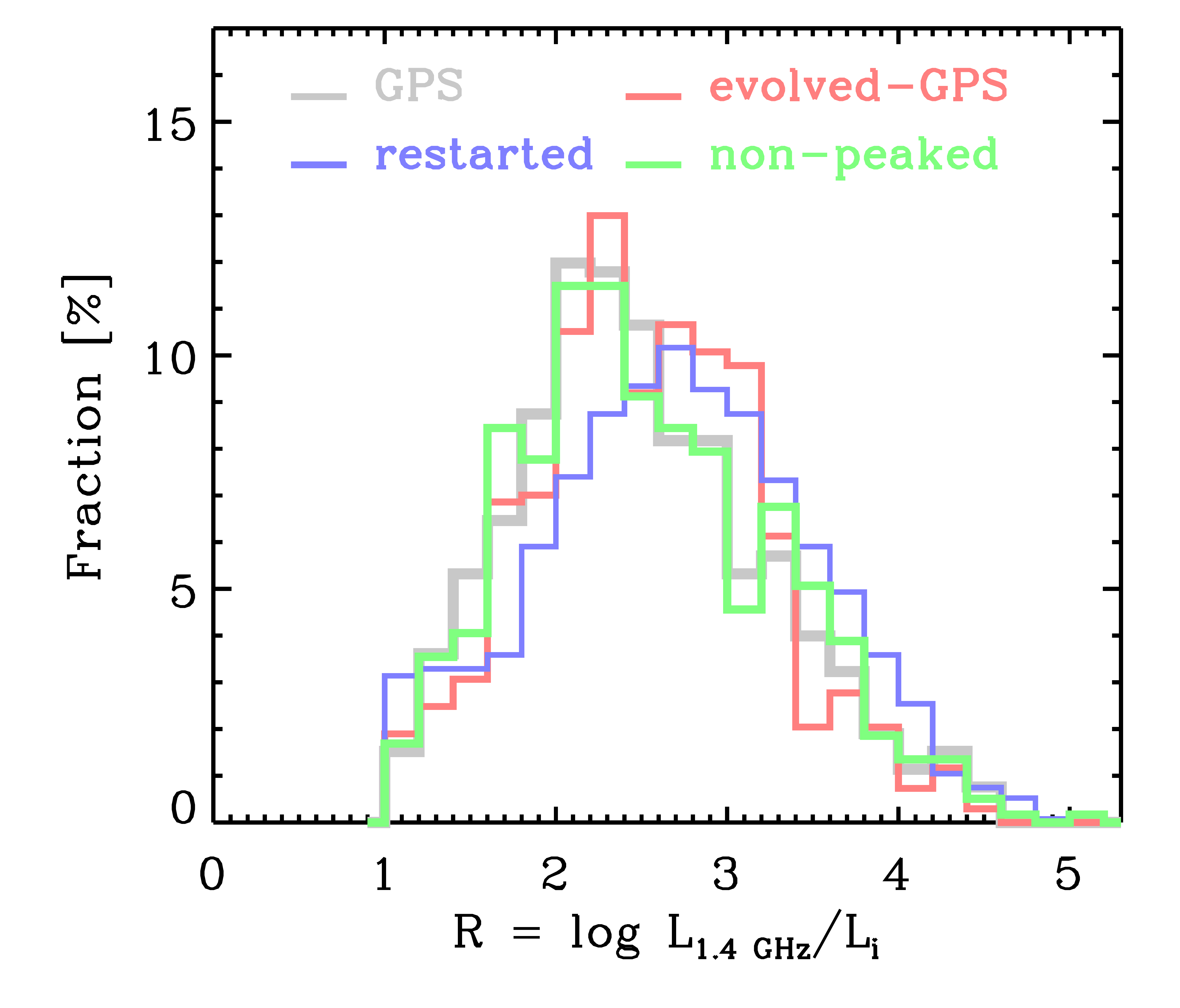}
\caption{The radio-loudness distributions. The light-gray, light-green, light-red, and the light-blue lines represent the GPS, evolved-GPS, restarted, and non-peaked quasars respectively.}
\label{fig:loundness2}
\end{figure}

\begin{figure}[ht!]
\includegraphics[width=0.48\textwidth]{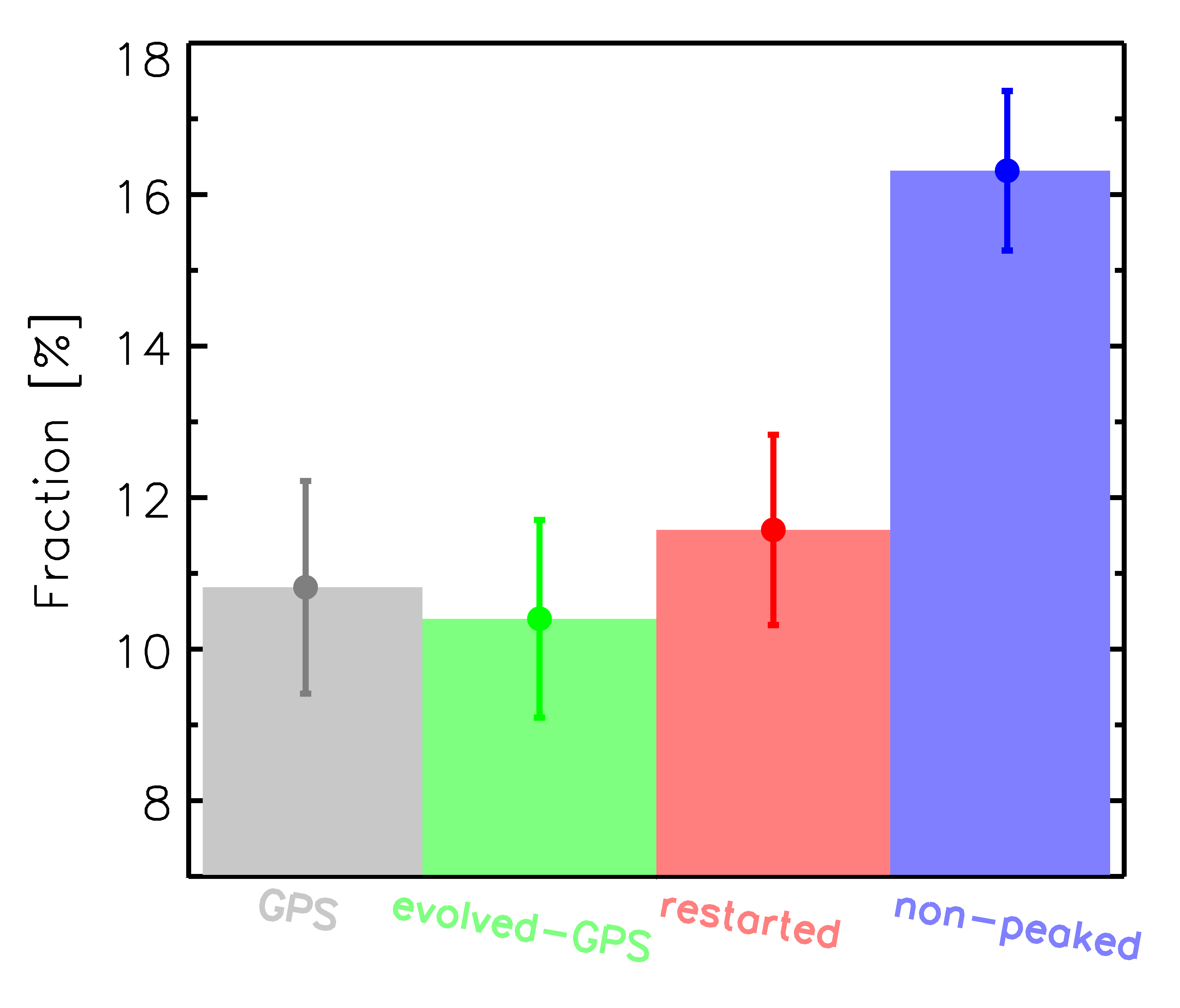}
\caption{The detection rates of \mgii\ associated absorption lines for the quasars with $R>1.5$. The light-gray, light-green, light-red, and the light-blue symbols represent the GPS, evolved-GPS, restarted, and non-peaked quasars, respectively.}
\label{fig:fr_abs2}
\end{figure}

Quasars exhibiting \mgii\ associated absorption lines typically display a redder spectrum compared to the general quasar population \citep[e.g.,][]{2012ApJ...748..131S,2020ApJ...893...25C,2022NatAs...6..339C}. This spectral characteristic could result in a lower $i$-band luminosity for quasars with \mgii\ associated absorption lines, potentially leading a subset of radio-quiet quasars to be categorized within the radio-loud population, particularly among the non-peaked quasars. For all quasar types, a significant majority of sources exhibit a radio-loudness substantially greater than 1 (Figure \textcolor{blue}{\ref{fig:loundness2}}). To further investigate the contamination from radio-quiet quasars, we focus on the radio-loud quasars with $R>1.5$ and reevaluate the incidence of \mgii\ associated absorption lines.  Non-peaked quasars have a noticeably higher $f_{c}$ compared to the other categories (Figure \textcolor{blue}{\ref{fig:fr_abs2}}). Consequently, the reddening effect caused by dust within the absorbing clouds does not significantly influence our primary findings.

\begin{figure}[ht!]
\includegraphics[width=0.48\textwidth]{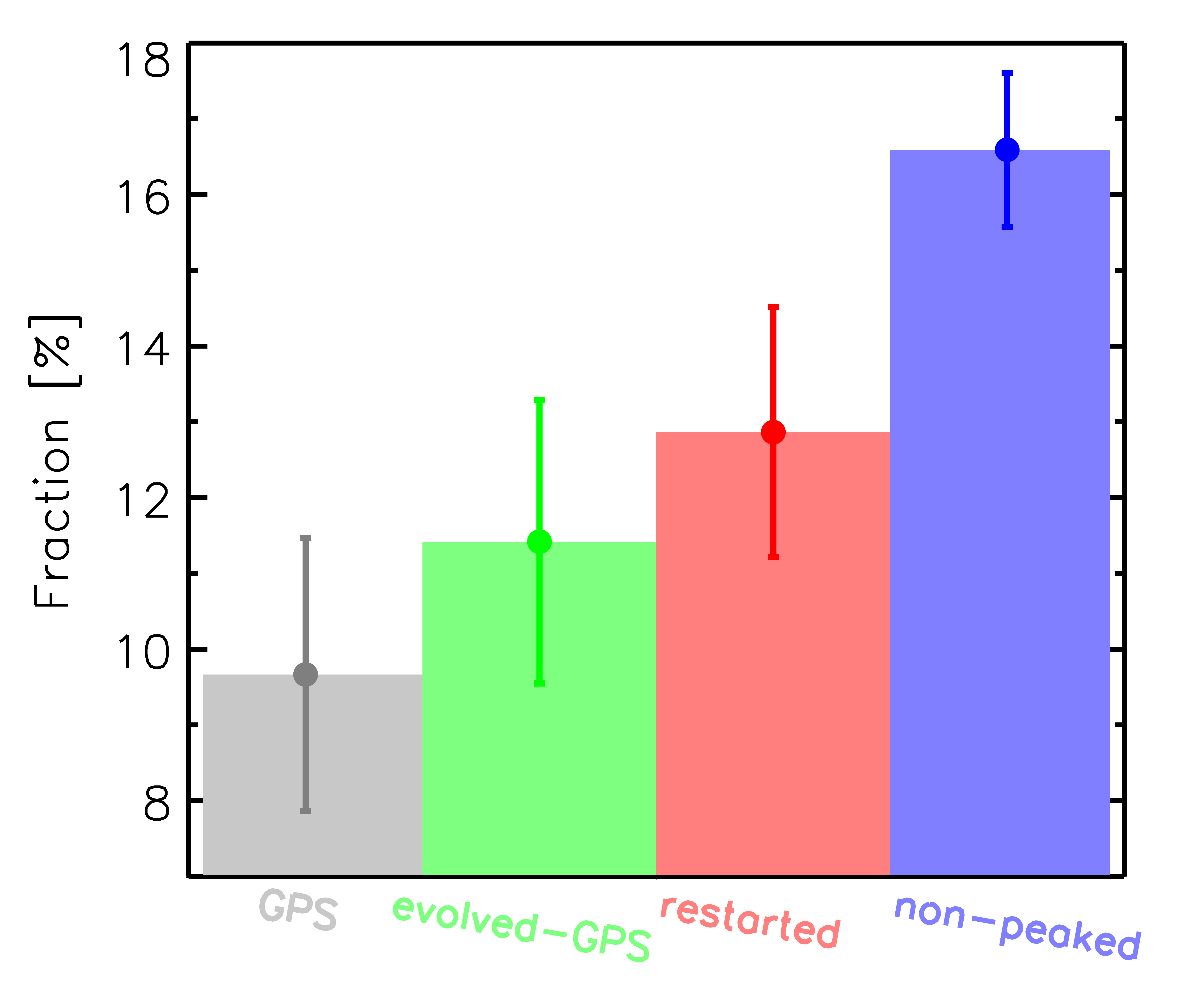}
\caption{The detection rates of \mgii\ associated absorption lines for the quasars with $|\alpha^{\rm 144}_{\rm 1400}| > 0.5$ and $|\alpha^{\rm 1400}_{\rm 3000}| > 0.5$. The light-gray, light-green, light-red, and the light-blue symbols represent the GPS, evolved-GPS, restarted, and non-peaked quasars, respectively.}
\label{fig:fr_abs3}
\end{figure}

There is currently a lack of consensus in defining flat-spectrum quasars. In this study, we characterize flat-spectrum quasars as those with $-0.5 \le \alpha^{\rm 144}_{\rm 1400} < 0$ and $-0.5 \le \alpha^{\rm 1400}_{\rm 3000} < 0$, which is similar to the definition used by \cite{2023A&A...674A.198K}. To assess the impact of our definition of flat-spectrum quasars, we consistently identify flat-spectrum sources with $|\alpha^{\rm 144}_{\rm 1400}| < 0.5$ and $|\alpha^{\rm 1400}_{\rm 3000}| < 0.5$ in each quadrant, following the approach outlined by \cite{2017AN....338..700M}. Upon excluding these flat-spectrum quasars, we present the incident rates of \mgii\ associated absorption lines in Figure \textcolor{blue}{\ref{fig:fr_abs3}}. It is apparent that both Figures \textcolor{blue}{\ref{fig:fr_abs}} and \textcolor{blue}{\ref{fig:fr_abs3}} demonstrate a similar distribution of $f_{c}$. This observation indicates that the selection criteria for flat-spectrum quasars does not significantly alter our primary conclusions.

The radio spectral shapes of quasars serve as effective indicators of their evolutionary state. In this context, GPS quasars represent the earliest stage, while non-peaked quasars are indicative of the most evolved sources. The specific evolutionary stages of GPS, evolved-GPS, and restarted quasars can be complex, and distinguishing their precise phase of development poses challenges. The evolutionary trajectory of quasars likely follows the sequence: GPS $\rightarrow$ evolved-GPS $\rightarrow$ restarted $\rightarrow$ non-peaked. The AGN feedback mechanism plays a crucial role in regulating gas distributions external to host galaxies throughout the source's lifetime \citep[e.g.,][]{2020ApJ...904....8C,2020ApJ...898...26G,2024A&A...689A.193A}. The observed evolutionary behavior in the incidence rate of \mgii\ associated absorption lines may relate to the distribution of gas within the host galaxy or CGM, which evolves as quasars mature and their jets expand. During the initial evolutionary phase, quasars and jets may be less than $10^5$ --- $10^6$ years old \citep[e.g.,][and references therein]{1998PASP..110..493O,2016AN....337....9O,2017ApJ...836..174C,2021A&ARv..29....3O,2022A&A...668A.186S}, exhibiting small-scale jets. These jets thus are less likely to eject substantial amounts of gas from the central region of its host to larger scales, resulting in a limited coverage fraction of absorbing gas for young quasars. As quasars progress to more mature or later evolutionary stages, typically around $10^8$ years \citep[e.g.,][]{2013MNRAS.435.3353H,2017MNRAS.469..639H,2017NatAs...1..596M,2019A&A...622A..12H,2023MNRAS.523..620P}, their jets have sufficient time to expand and eject significant quantities of gas from the center to more distant regions, including the CGM of the host galaxy. This leads to increased coverage fraction of absorbing gas. Therefore, the incidence rate of \mgii\ absorption lines is observed to rise significantly as quasars transition from their young stages to older ages, representing a natural consequence of the feedback processes originating from these quasars.

\section{Summary} \label{sec:summary}
We present a compilation of 3,141 radio-loud quasars within the redshift range $0.28 < z_{\rm em} < 2.6$, sourced from the surveys conducted by the SDSS, LoTSS, FIRST, and VLASS, deliberately excluding the flat-spectrum radio sources. Among these quasars, we identified 418 sources having SDSS optical spectra that exhibit \mgii\ associated absorption lines with velocity offset $\upsilon_{r} < 5000$ \kms. Utilizing radio observations, we calculated radio spectral indices ($\alpha^{\rm 1400}_{\rm 3000}$ between 1400 and 3000 MHz; $\alpha^{\rm 144}_{\rm 1400}$ between 144 and 1400 MHz) that aid in delineating the evolutionary trajectory of quasars, along the source sequence GPS $\rightarrow$ evolved-GPS $\rightarrow$ restarted $\rightarrow$ non-peaked.

We established a quantitative relationship between the evolutionary stage of quasars and the incidence rate ($f_{c}$) of \mgii\ associated absorption lines. The incidence rate significantly increases from $f_{c} = 0.10 \pm 0.01$ to $0.17 \pm 0.01$ as quasars progress from the GPS phase to the non-peaked phase. This observed trend is likely a result of jet-driven feedback, where jets expel substantial amounts of gas from small scales to larger scales within host galaxies, extending even to the scales of the CGM as quasars advance and their jets evolve. This mechanism effectively enhances the coverage fraction of absorbing gas.

\section*{Acknowledgements}
CZF is supported by the Guangxi Natural Science Foundation (2024GXNSFDA010069), the National Natural Science Foundation of China (12473011,12073007), and the Scientific Research Project of Guangxi University for Nationalities (2018KJQD01). LCH is supported by the National Key R\&D Program of China (2022YFF0503401), the National Science Foundation of China (11991052, 12233001), and the China Manned Space Project (CMS-CSST-2021-A04, CMS-CSST-2021-A06). We thank the anonymous referee for helpful suggestions.

%XX Many references are in the wrong format. e.g., why are the journal names spelled out in full?

\bibliography{cG}{}
\bibliographystyle{aasjournal}

%% This command is needed to show the entire author+affiliation list when
%% the collaboration and author truncation commands are used.  It has to
%% go at the end of the manuscript.
%\allauthors

%% Include this line if you are using the \added, \replaced, \deleted
%% commands to see a summary list of all changes at the end of the article.
%\listofchanges

\end{document}